\documentclass[conference]{IEEEtran}
\IEEEoverridecommandlockouts
\usepackage{cite}
\usepackage{amsmath,amssymb,amsfonts}
\usepackage{algorithmic}
\usepackage{graphicx}
\usepackage{textcomp}
\usepackage{xcolor}
\def\BibTeX{{\rm B\kern-.05em{\sc i\kern-.025em b}\kern-.08em
    T\kern-.1667em\lower.7ex\hbox{E}\kern-.125emX}}
\begin{document}

\title{Spectrum Occupancy Detection Supported by Federated Learning

\thanks{Copyright © 2023 IEEE. Personal use is permitted. For any other purposes, permission must be obtained from the IEEE by emailing pubspermissions@ieee.org. This is the author’s version of an article that has been published in the proceedings of 2023 International Conference on Software, Telecommunications and Computer Networks (SoftCOM) and published by IEEE. Changes were made to this version by the publisher prior to publication, the final version of record is available at: https://ieeexplore.ieee.org/document/10271668. To cite the paper use: Ł. Kułacz, "Spectrum Occupancy Detection Supported by Federated Learning," 2023 International Conference on Software, Telecommunications and Computer Networks (SoftCOM), Split, Croatia, 2023, pp. 1-3, doi: 10.23919/SoftCOM58365.2023.10271668.
The presented work has been funded by the National Science Centre in Poland within the project (no. 2021/41/N/ST7/01298) of the PRELUDIUM program.}

}

\author{\IEEEauthorblockN{\L. Ku\l acz }
\IEEEauthorblockA{\textit{Institude of Radiocommunication} \\
\textit{Poznan University of Technology}\\
Poznan, Poland \\
lukasz.kulacz@put.poznan.pl}
}

\maketitle

\begin{abstract}
Dynamic spectrum access is essential for radiocommunication and its limited spectrum resources. The key element of dynamic spectrum access systems is effective spectrum occupancy detection. In many cases, machine learning algorithms improve detection effectiveness. Because of the recent trend of using federated learning, a federated learning algorithm is presented in the context of distributed spectrum occupancy detection. The results of the work presented in the paper are based on actual signal samples collected in the laboratory. The~proposed algorithm is effective, especially in the context of a set of sensors with faulty sensors.
\end{abstract}


\section{Introduction}

Given the enormous pace of development of wireless technologies and the problem of insufficient spectrum resources, dynamic spectrum access is a very promising solution to meet the requirements for radiocommunication systems. The possibility of using the spectrum for purposes other than those for which it was originally intended (reserved) seems to be the easiest conceptual solution. Unfortunately, for many reasons, this solution is not so straightforward. First, a Dynamic Spectrum Access (DSA) system is required, which will be able, using appropriate algorithms and data, to ensure access to the spectrum for secondary (additional) users~\cite{b_dsa}. At the same time, care should be taken to ensure that the primary user's transmission is provided with the best possible quality of service, whether through appropriate separation of resources between users, control of the transmission power of system users, or even temporary disabling the possibility of transmission for unlicensed users. The critical component of the indicated system is access to information about the presence of primary users' transmissions. Once the system has this information, it can appropriately protect the primary user's transmission. 
The primary source of information on spectrum occupancy is spectrum sensing, a huge and separate part of the radiocommunication research area. Many papers deal with spectrum occupancy detection algorithms in various scenarios~\cite{b_fed}. Machine learning algorithms are also successfully used to improve the detection quality, which supports the detection process by extracting essential, but not always noticeable, details of the transmission~\cite{b_ml}. An example of such details may be the periodicity of transmission, different traffic volumes depending on the time of day or year, or the characteristics of the transmission system used. It is worth noting one of the significant features of wireless systems related to the use of machine learning to detect spectrum occupancy - a very rare situation is the ability to obtain verified information about whether a signal is transmitted in a given place and time. This is the most critical aspect of supervised machine learning and the collection of training data necessary for its operation. In addition, it was noted that knowledge from a~single sensor could often be distorted due to the sensor's specific location or its unreliability. To eliminate this problem, many cooperative spectrum sensing algorithms have been considered. They allow us to make better (more reliable) decisions in a specific area; however, using machine learning algorithms is more complicated due to the massive amount of data necessary to be transferred between the sensors and the collecting node. This reduces the time available for the actual transmission of user data and thus reduces the system's efficiency, ultimately aiming to send as much user data as possible. It should be remembered that sensors should also be straightforward devices, which is an~additional challenge. Another critical problem is the transparency of the transmitted data, i.e., the transmission of the collected measurement data by sensors can be relatively easily intercepted and analyzed and, worse, replaced~\cite{b1}.
One of the promising ideas for solving this problem, which has been gaining popularity recently and is widely considered in many aspects and applications, is federated learning~\cite{b2}. 

\section{Federated Learning}

Federated learning assumes the operation of individual sensors in a specific area of the network, which collects data in their environment while training their machine learning model. Then, the model coefficients of individual sensors are exchanged, as opposed to the samples collected in cooperative spectrum sensing, and an aggregated model is created (for example, in the simplest Federated Averaging (FedAvg) scenario - created by averaging all coefficients received from the sensors)~\cite{b3}. Further, the collective model, and its coefficients, are transferred to all sensors. In the simplest scenario, the sensors use a~collective model. Then, based on the newly collected samples, they slightly adjust its coefficients or create a different model resulting from the appropriate merge of the collective model and their own model. The procedure described above is repeated many times. Consequently, individual sensors have access to a collective model, which is at least indirectly based on many more samples than those collected by a single sensor. In addition, the amount of information transferred between the sensors is limited to only the model coefficients, which depends on the complexity of the machine learning model~\cite{b4}. One benefit of using federated learning is the ability to train a~single-sensor model much faster in case of its replacement or simply adding a new one. Therefore, the speed of adaptation of such a system should be its advantage - especially in the context of the aforementioned general challenge related to spectrum occupancy detection. Similarly, the ability to detect a faulty sensor from ''significantly'' different reported model coefficients alone should be a valuable safeguard for the system - increasing its reliability. Federated learning used in this way cannot offer higher detection efficiency than cooperative spectrum occupancy detection due to limited access to training data (only indirectly). However, at the expense of slightly lower detection efficiency, it is possible to significantly reduce the amount of sent control data and improve system security~\cite{b5}.

\section{Data Collection}

The following measurements were carried out to demonstrate the operation of the spectrum occupancy detection system using federated learning. Two sets of devices were used, each consisting of a Linux computer with GNU Radio software (version 3.8) and a Universal Software Radio Peripheral (USRP) B210 device, where one set acts as a transmitter and the other as a receiver of the test radio signal. The transmitted signal was a random set of values modulated with Gaussian Minimum Shift Keying (GMSK). The receiver recorded a specified number of signal samples (IQ) for successively higher transmit signal gain settings. The receiver collected 100 million IQ samples for noise and 10 million IQ samples per transmit power level used by the transmitter. The center frequency was set to 2.1 GHz and the bandwidth to 40 MHz. In~addition, the receiver collected reference data as noise samples when the transmitter was turned off. The received signal was divided into every 10'000 samples from which Fast Fourier Transform (FFT) was calculated. Then the average power received in this channel was calculated for each channel and the kurtosis and skewness of the autocorrelation function. In addition, a label has been adapted to each set of these values depending on whether the transmitter is turned on during a~given measurement. The actual data collected in the described manner was then used in a spectrum occupancy detection simulation using federated learning. For simplicity, the current simulation results use data for only one frequency channel (also, the tested signal was transmitted only in one of the analyzed channels). The data presented in this study are openly available~\cite{data_source}.

\section{Simulation Setup and Reference Scenario}

The first step in preparing data for machine learning purposes was to balance the number of measurements for each label. Because after the measurements, most of the data was collected for the signal's turned-on transmission- some measurements with significant transmit power were omitted. In this way, the same amount of noise and signal data (with a transmit power close to the noise power) were used to train the machine learning model. Since the experiment is about federated learning, all the collected data was shuffled and divided into five equal subsets - simulating that each of the five sensors contained some stored information. This is a~deliberate simplification to focus primarily on analyzing the federated learning algorithm.
The effectiveness of determining the spectrum occupancy was calculated (for each sensor separately) based on the classic energy detection algorithm (assuming a false alarm probability of 1\%) to have a~reference point for all obtained measurements. The average efficiency of this algorithm is 93.18\%, whereby efficiency means the percentage of correctly determined spectrum occupancy values in the set of test values (all collected data). The spectrum occupancy detection system's efficiency is measured by the accuracy of the ML model used in this process (efficiency and accuracy are used alternately). Then, a simple machine learning algorithm, logistic regression, was tested. In this case, the average efficiency of spectrum occupancy determination was 98.83\%.  In addition, a neural network was tested (with an input layer with four nodes, a~hidden layer with four nodes, and an output layer with one node), achieving an average of 99.04\% efficiency. Data from each sensor was split into training (80\%) and testing (20\%) datasets. Each of the mentioned algorithms has been tested for many different parameters and validated using the stratified K-Fold cross-validation algorithm; however, the effectiveness base on the entire dataset (from the perspective of spectrum occupancy detection) has been shown to standardize the results. 


\section{Simulation Results}

\begin{figure}[htbp]
\centerline{\includegraphics[width=0.40\textwidth]{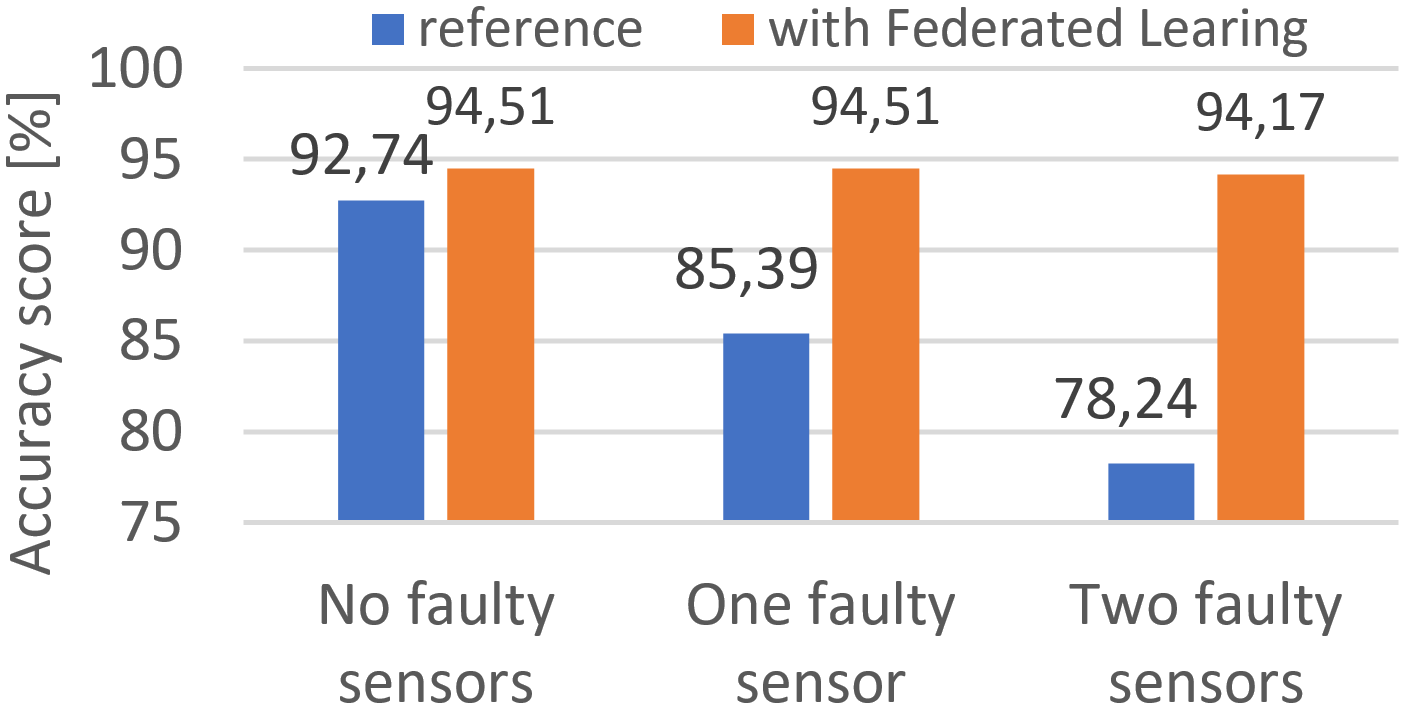}}
\caption{Accuracy score for Logistic Regression machine learning model utilizing all collected data}
\label{fig2}
\end{figure}

The conducted simulation of the operation of the federated learning algorithm consisted in dividing the training data into equal batches and iteratively: training single models (for each sensor separately - using only training data for a~single sensor), then creating a collective model based on the average value of model coefficients and setting the calculated coefficients in the models of each sensor. In parallel, copies of the models for each sensor were retained (omitted from the coefficient exchange process) to reference the scenario without using federated learning. First, federated learning using a logistic regression model was tested. The average efficiency of the presented federated learning algorithm was 94.51\%, whereas the average efficiency of the sensors (without federated learning) was 92.74\%. It is worth explaining the differences between the obtained values and the aforementioned efficiency of the logistic regression algorithm itself (98.83\%). This difference is since in federated learning, training data enters the model gradually, and after obtaining only a small part of the data, we try to estimate the effectiveness of the model. In the reference scenario, however, all data is transferred to the model and subjected to many iterations of the learning process. More interesting results can be seen when one of the analyzed sensors is faulty and reports random decisions about spectrum occupancy. In this situation, the average efficiency of the presented federated learning algorithm was 94.51\%, whereas the average efficiency of the sensors (without federated learning) was 85.39\%. With two faulty (in the same way) sensors, the average efficiency of the presented federated learning algorithm was 94.17\%, whereas the average efficiency of the sensors (excluding federated learning) was 78.24\%. It can be noticed that using even such a simple algorithm as averaging the model coefficients improves the system's reliability when simple and unreliable sensors are used. Fig.~\ref{fig2} shows the average accuracy of predictions on the analyzed dataset in the scenario with Logistic Regression used in the FL process.

\begin{figure}[htbp]
\centerline{\includegraphics[width=0.40\textwidth]{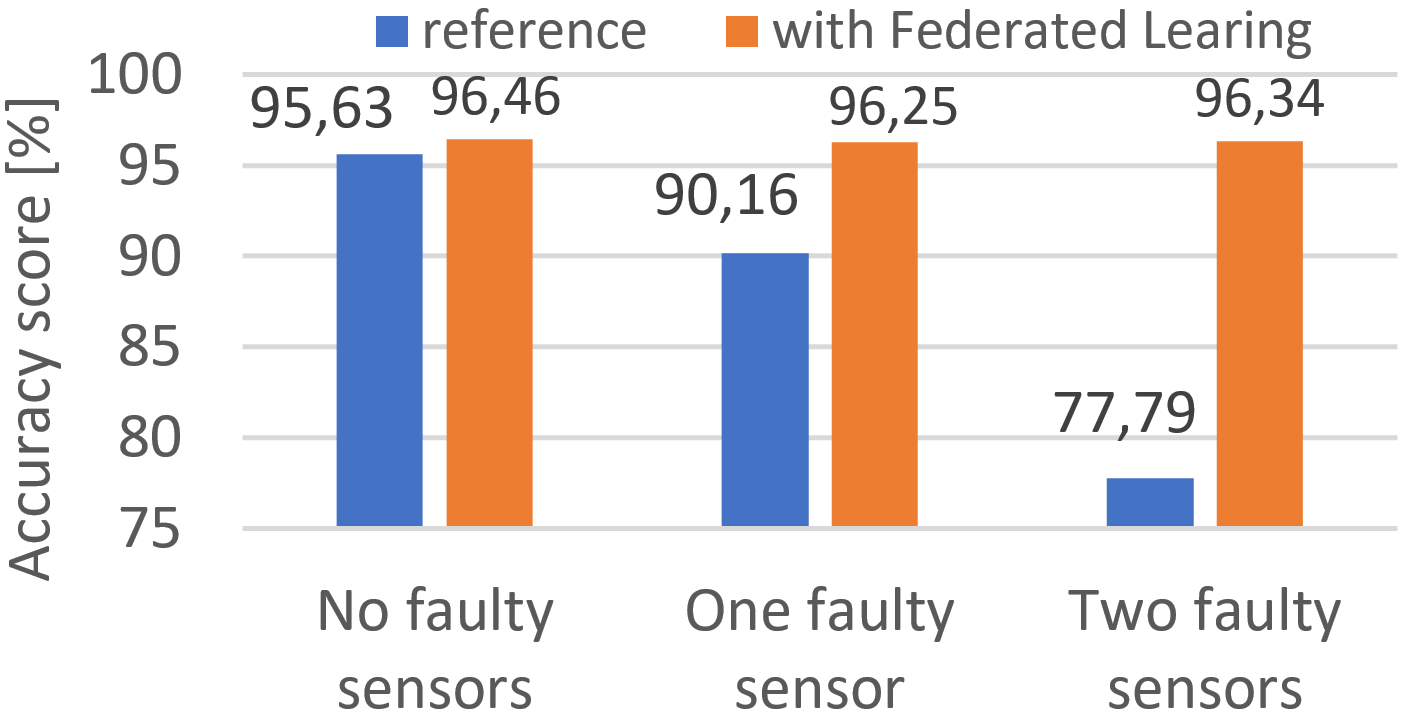}}
\caption{Accuracy score for Neural Network machine learning model utilizing all collected data}
\label{fig3}
\end{figure}

Then, federated learning using a neural network was tested. The average efficiency of the presented federated learning algorithm was 96.46\%, whereas the average efficiency of the sensors (excluding federated learning) was 95.63\%. As before, the difference with reference results is that training data enters the model gradually in federated learning. After obtaining only a small part of the data, we try to estimate the model's effectiveness. In the reference scenario, however, all data is transferred to the model and subjected to many iterations of the learning process. Here, the scenario where one of the analyzed sensors is faulty, and reports random decisions about spectrum occupancy was checked again. In this situation, the average efficiency of the presented federated learning algorithm was 96.25\%, whereas the average efficiency of the sensors (excluding federated learning) was 90.16\%. With two broken (in the same way) sensors, the average efficiency of the presented federated learning algorithm was 96.34\%, whereas the average efficiency of the sensors (excluding federated learning) was 77.79\%. It can be noticed that using even such a simple algorithm as averaging model coefficients, also in the case of neural networks, improves system reliability when simple and unreliable sensors are used. Fig.~\ref{fig3} shows the average accuracy of predictions on the analyzed data set in the scenario with Logistic Regression used in the Federated Learning process.  However, using neural networks in such a~simple scenario requires exchanging a much larger number of coefficients (41 instead of 4), and the model-learning process is computationally more complex than logistic regression. It is also worth emphasizing that it is possible to detect an incorrect model in both of the analyzed situations based on how much its coefficients differ from those of other models.

\section{Conclusions}
The algorithms presented in the work show potential for using federated learning in spectrum occupancy detection systems. In specific scenarios, they provide increased reliability of detection and, more importantly, ensure greater efficiency of decisions made compared to classic algorithms (such as energy detection) and comparable efficiency compared to algorithms using machine learning (compared to a model trained on a~larger number of samples - such as in cooperative spectrum occupancy detection). Further work on this topic is required, mainly to investigate the influence of the environment change on the created models, especially when received signal power in the sensors will be close to the noise level. Moreover, simplification made in this work, i.e., collection of samples only by one USRP device, should be extended - as hardware imperfections, even minor differences in received signal level, could be critical in federated learning.


\begin{thebibliography}{00}

\bibitem{b_dsa} T. Dudda, and T. Irnich, ''Capacity of cellular networks deployed in TV White Space,''
IEEE International Symposium on Dynamic Spectrum Access Networks, pp. 254–265, 2012.

\bibitem{b_fed} M. Wasilewska, H. Bogucka, and A. Kliks, ''Federated Learning for 5G Radio Spectrum Sensing,'' Sensors, vol. 22, no. 1, 2021.

\bibitem{b_ml} V. P. Rekkas, et al. ''Machine Learning in beyond 5G/6G networks—state-of-the-art and future trends,'' Electronics, vol. 10, no. 22, p. 2786,2021.

\bibitem{b1} T. Yucek, H. Arslan, ''A survey of spectrum sensing algorithms for cognitive radio applications,'' IEEE Communications Surveys Tutorials, vol. 11, no. 1, pp. 116–130, 2009.
\bibitem{b2} J. Tian, P. Cheng, Z. Chen, M. Li, and H. Hu, ''A Machine Learning-Enabled Spectrum Sensing Method for OFDM Systems,'' IEEE Transactions on Vehicular Technology, vol. 68, no. 11, Nov. 2019. 
\bibitem{b3} M. Troglia, J. Melcher, Y. Zheng; D. Anthony, A. Yang, T. Yang, ''FaIR: Federated Incumbent Detection in CBRS Band,'' 2019 IEEE International Symposium on Dynamic Spectrum Access Networks (DySPAN), Newark, NJ, USA, 19 December 2019.
\bibitem{b4} S. Abdulrahman, H. Tout, H. Ould-Slimane, A. Mourad, C. Talhi, M. Guizani. ''A Survey on Federated Learning: The Journey from Centralized to Distributed On-Site Learning and Beyond,'' IEEE Internet of Things Journal, vol. 8, no. 7, April 2021, pp. 5476 – 5497.
\bibitem{b5} C. Campolo, G. Genovese, G. Singh, A. Molinaro, ''Scalable and interoperable edge-based federated learning in IoT contexts, '' Computer Networks, vol. 223, 2023.

\bibitem{data_source} \L . Kułacz, ''GMSK Spectrum Sensing for Machine Learning''. Zenodo, Oct. 19, 2023. doi: 10.5281/zenodo.10021228.




\end{thebibliography}
\end{document}